\documentclass[11pt]{article}
\begin{document}

\title{ A note on Non-commutativity and the Zero Point Field}

\author{B.G. Sidharth\footnote{birlasc@gmail.com}, B.M. Birla Science Centre,\\ Adarsh Nagar, Hyderabad - 500 063, India\\
\\Abhishek Das\footnote{parbihtih3@gmail.com}, B.M. Birla Science Centre,\\ Adarsh Nagar, Hyderabad - 500 063, India\\
\\Arka Dev Roy\footnote{arkaroyrph@gmail.com}, B.M. Birla Science Centre,\\ Adarsh Nagar, Hyderabad - 500 063, India\\}
\maketitle

%%%%%%%%%%%%%%%%%%%%%%%%%%%%%%%%%%%%%%%%%%%%%%%%%%%%%%%%%%%%%%%%%%%%%%%%%%%%%%%%%%%%%%%%%%%%
\begin{abstract}
In this paper we endeavour to find a connection between the non-commutative nature of space time and the {\it zero point field}. We observe that extra effects come into play when we take into account the Compton scale effects in such a space-time and the electromagnetic field tensor and the current density get modified. This defines an underlying connection between non-commutativity and the {\it zero point field}.
\end{abstract}
\maketitle

In contradistinction to the usual commutative space-time, a non-commutative space-time has been considered and studied by many authors including Sidharth \cite{bgs}, as the framework of various fundamental phenomena. This non-commutative nature of space-time was investigated by H. Snyder in the context of the infrared catastrophe of soft photons in the Compton scattering and in general to renormalize quantum field theory by applying the noncommutative quantized space-time. Besides, such a framework has been widely studied by other authors \cite{Battisti,Meljanac,Chaichian,Mendes,Camelia} also. We begin with the following fundamental relation of non-commutativity \cite{Snyder} as\\
\begin{equation}
[x_{\mu}, x_{\nu}] = \epsilon \beta(l^2)
\end{equation}
where, $\epsilon$ is a constant and $\beta(l^2)$ a suitable matrix, is some linear function of the square of the Compton length ($l$) of the electron which is very small. Here, $l$ is the minimal physical length. Normally in modern Quantum Gravity approaches the minimum length $l$ is taken to be the Planck length \cite{a,b}. But over the years we have considered $l$ to be the Compton wavelength \cite{c,d,db,e}. To see how this works let us go back to the Dirac coordinate \cite{diracpqm}
\begin{equation}
x = (c^2p_1H^{-1}t) + \frac{\imath}{2} c\hbar (\alpha_1 -
cp_1H^{-1})H^{-1}\label{2.22}
\end{equation}
with similar expressions for the other two coordinates. The first term in (\ref{2.22}) gives the usual position coordinates which commute with one another. It is the second term which gives the Zitterbewegung spread over the Compton wavelength. Dirac himself explained that this rapidly oscillating term would not pose a difficulty because physical measurements are spread over a small spacetime interval of the order of the Compton scale, an average over which would remove the term. In other words our usual spacetime coordinates are really values averages over the Compton wavelength \cite{diracpqm}. Let us analyse (\ref{2.22}) in a little greater detail. Let us write it as
\begin{equation}
x_l= \bar{x}_l+ \Theta_{lk} \bar{p}_k,
\end{equation}
\begin{equation}
p_l = \bar{p}_l,
\end{equation}
where the $x$s, $\bar{x} \, \mbox{and}\, \bar{p}$ obey the usual commutation relations.
\begin{equation}
[\bar{x}_l , \bar{x}_k] = 0,
\end{equation}
\begin{equation}
[\bar{x}_l , \bar{p}_k] = \imath \hbar \delta_{lk},
\end{equation}
\begin{equation}
[\bar{p}_l , \bar{p}_k] = 0.
\end{equation}
$\bar{x}$ represents the averaged space coordinate, that is the first term in (\ref{2.22}). However it is easy to verify that the $x$s and $p$s, as can be easily verified from equations (2) and (3) satisfy
\begin{equation}
[x_l , x_k] = -2\imath \hbar \Theta_{lk},
\end{equation}
\begin{equation}
[x_l , p_k] = \imath \hbar \delta_{lk},
\end{equation}
\begin{equation}
[p_l , p_k] = 0,
\end{equation}
This is precisely the equation (1) referred to above, which alternatively follows from the Snyder treatment if the Compton wavelength is taken as the minimum length. It may be further pointed out that as noted by Wigner and Salecker \cite{wigner} there can be no physical measurements within the Compton wavelength.\\
At this point we would like to emphasize the well known close relationship between the Zero Point Field, which manifests itself even when no external fields are applied and zitterbewegung which as Schrodinger first noticed -- and this can be attributed to the Zero Point fluctuations \cite{bd,calphys}. Indeed all of space is filled with the Zero Point Field (Cf.ref.\cite{calphys}).\\
In fact this was the basis for Sidharth's 1997 cosmology, which correctly predicted that the universe would be accelerating with a small cosmological constant, at a time when the ruling paradigm was exactly the opposite.\\
Now, with relation (1) we will also have the following relation for differentials as \\
\begin{equation}
[{\rm d}x_{\mu}, {\rm d}x_{\nu}] = \epsilon \beta(l^2)
\end{equation}
since, the differentials too are small lengths or coordinates. Therefore, we have\\
\[{\rm d}x_{\mu}{\rm d}x_{\nu} = {\rm d}x_{\nu}{\rm d}x_{\mu} + \epsilon \beta(l^2)\]\\
From this relation we may write\\
\[\frac{{\rm d}^2}{{\rm d}x_{\mu}{\rm d}x_{\nu}} = \frac{{\rm d}^2}{{\rm d}x_{\nu}{\rm d}x_{\mu}}[\frac{1}{1 + \frac{\epsilon \beta(l^2)}{{\rm d}x_{\nu}{\rm d}x_{\mu}}}]\]\\

Now, since $\beta(l^2)$ is very small the above relation can be approximated as\\
\begin{equation}
\frac{{\rm d}^2}{{\rm d}x_{\mu}{\rm d}x_{\nu}} = \frac{{\rm d}^2}{{\rm d}x_{\nu}{\rm d}x_{\mu}}[1 - \frac{\epsilon \beta(l^2)}{{\rm d}x_{\nu}{\rm d}x_{\mu}}]
\end{equation}
In terms of partial derivatives we can write\\
\begin{equation}
\frac{\partial^2}{\partial{x_{\mu}}\partial{x_{\nu}}} = \frac{\partial^2}{\partial{x_{\nu}}\partial{x_{\mu}}}[1 - \frac{\epsilon \beta(l^2)}{\partial{x_{\nu}}\partial{x_{\mu}}}]
\end{equation}
Now, let us consider the antisymmetric field tensor in the electromagnetic case as\\
\begin{equation}
F_{\mu\nu} = - F_{\nu\mu}
\end{equation}
Now, we know that this field tensor satisfies the relation\\
\[\frac{\partial^{2}F_{\mu\nu}}{{\partial{x_{\mu}}\partial{x_{\nu}}}} = 0\]\\
in the simple electromagnetic case. Here, the left hand side can also be written as\\
\[\frac{\partial^{2}F_{\mu\nu}}{{\partial{x_{\mu}}\partial{x_{\nu}}}} = \frac{1}{2}[\frac{\partial^{2}F_{\mu\nu}}{{\partial{x_{\mu}}\partial{x_{\nu}}}} - \frac{\partial^{2}F_{\nu\mu}}{{\partial{x_{\mu}}\partial{x_{\nu}}}}]\]\\
which gives (upon interchanging the indices $\mu$ and $\nu$ in the second term on the right hand side)\\
\[\frac{\partial^{2}F_{\mu\nu}}{{\partial{x_{\mu}}\partial{x_{\nu}}}} = \frac{1}{2}[\frac{\partial^{2}F_{\mu\nu}}{{\partial{x_{\mu}}\partial{x_{\nu}}}} - \frac{\partial^{2}F_{\mu\nu}}{{\partial{x_{\nu}}\partial{x_{\mu}}}}]\]\\
Now, from relation (13) we have\\
\[\frac{\partial^{2}F_{\mu\nu}}{{\partial{x_{\mu}}\partial{x_{\nu}}}} = \frac{1}{2}[\frac{\partial^{2}F_{\mu\nu}}{{\partial{x_{\mu}}\partial{x_{\nu}}}} - \frac{\partial^{2}F_{\mu\nu}}{{\partial{x_{\nu}}\partial{x_{\mu}}}}] = -\frac{1}{2}[\frac{\epsilon \beta(l^2)}{\partial{x_{\nu}}\partial{x_{\mu}}}]\frac{\partial^{2}F_{\mu\nu}}{{\partial{x_{\nu}}\partial{x_{\mu}}}}\]\\
Using relation (14) this yields
\begin{equation}
\frac{\partial^{2}F_{\mu\nu}}{{\partial{x_{\mu}}\partial{x_{\nu}}}} = \frac{1}{2}[\frac{\epsilon \beta(l^2)}{\partial{x_{\nu}}\partial{x_{\mu}}}]\frac{\partial^{2}F_{\nu\mu}}{{\partial{x_{\nu}}\partial{x_{\mu}}}}
\end{equation}

where, the right hand side arises due to the vacuum fluctuations of the electromagnetic field in a noncommutative space-time. Now, interchanging $\mu$ and $\nu$ ($\mu \leftrightarrow \nu$) on the right hand side of the above equation and writing \\

\[\frac{1}{2}[\frac{\beta(l^2)}{\partial{x_{\mu}}\partial{x_{\nu}}}]\frac{\partial^{2}F_{\mu\nu}}{{\partial{x_{\mu}}\partial{x_{\nu}}}} = \frac{\partial^{2}(F_{\mu\nu})_{0}}{{\partial{x_{\mu}}\partial{x_{\nu}}}}\]\\
where $(F_{\mu\nu})_{0}$ is the field tensor for the {\it zero point field}, we get\\
\begin{equation}
\frac{\partial^{2}F_{\mu\nu}^{\prime}}{{\partial{x_{\mu}}\partial{x_{\nu}}}} = 0
\end{equation}
or,
\begin{equation}
\frac{\partial j_{\mu}^{\prime}}{\partial x_{\mu}} = 0
\end{equation}
where,
\begin{equation}
F_{\mu\nu}^{\prime} = F_{\mu\nu} - \epsilon (F_{\mu\nu})_{0}
\end{equation}
and
\begin{equation}
j_{\mu}^{\prime} = j_{\mu} - \epsilon (j_{\mu})_{0}
\end{equation}
are respectively the total field tensor and the total current density respectively. We should remember that the last two relations are comprised of the normal electromagnetic field and the {\it zero point field} or the field of the quantum vacuum. Averaging over these fluctuations\\
\[\frac{1}{2}[\frac{\beta(l^2)}{\partial{x_{\mu}}\partial{x_{\nu}}}]\frac{\partial^{2}F_{\mu\nu}}{{\partial{x_{\mu}}\partial{x_{\nu}}}} = \frac{\partial^{2}(F_{\mu\nu})_{0}}{{\partial{x_{\mu}}\partial{x_{\nu}}}}\]\\
up to the Compton scale one would obtain the total contribution from the {\it zero point field}.\\

Now, let us consider the relations (18) and (19) as identifications. These two equations will modify the Maxwell equations of electromagnetism owing to the modified field tensor $F_{\mu\nu}^{\prime}$. On the other hand, in a previous paper \cite{bgs2} we had considered the following identification based on the presence of the {\it zero point field}\\
\begin{equation}
F^{\prime}_{\mu \nu} = (F_{\mu \nu})_0 + \xi (F_{\mu \nu})
\end{equation}
corresponding to a modified vector potential \\
\[A^{\prime}_{\mu} = A_{\mu0} + \xi A_{\mu}\]\\
and a current density\\
\[j^{\prime}_{\mu} = j_{\mu0} + \xi j_{\mu}\]\\
where, $(A_\mu)_{0}$ is the {\it vector potential} associated with the {\it zpf}, $j_{\mu0}$ is the current density corresponding to $(A_\mu)_{0}$ and $\xi$ is a convenient constant which can be taken to be unity for the sake of simplicity. With these considerations it was shown that considering a modified field tensor as $F_{\mu\nu}^{\prime}$ in equation (20) the {\it anomalous} gyromagnetic ratio for the electron is explained very elegantly \cite{bgs2} including Schwinger's correction terms. In fact, it has been shown there that the current density $j_{\mu0}$ in the Compton scale is the reason for the anomaly in the {\it gyromagnetic ratio} of the electron.\\
Here, we have derived equation (18) which is analogous to equation (11) from a different standpoint of non-commutativity. Precisely, if we set $\epsilon = -1$ in (18) and $\xi = 1$ in (20) then we have the same equations. Thus, we are able to perceive a rationale for the relations (18) and (19) in the sense that these extra effects give rise to certain phenomena that occur due to {\it zitterbewegung} effects in the Compton scale. On the other hand, it is manifest from equations (18) and (19) that if $\epsilon = 0$ then we get back the usual covariant Maxwell's equations due to the normal field tensor $F_{\mu\nu}$.\\
In this manner we find that the non-commutative nature of space time and the {\it zero point field} is intrinsically connected in the Compton scale. It is natural that we get the extra effects in the relations (15), (16) and (17). Since, the Compton length $l$ and consequently $\beta(l^2)$ is extremely small it is easy to conceive that the extra effects are almost negligible. Therefore, it is almost natural that even if we neglect $\beta(l^2)$ or set $\epsilon = 0$ we get convenient results. But, it is undeniable that there are some extra effects when we are in the domain of Compton scale.\\
Now, we have a found a relation between the electromagnetic tensor ($F_{\mu\nu}$) and the {\it zpf} tensor $(F_{\mu\nu})_{0}$ as\\
\begin{equation}
\frac{1}{2}[\frac{\beta(l^2)}{\partial{x_{\mu}}\partial{x_{\nu}}}]\frac{\partial^{2}F_{\mu\nu}}{{\partial{x_{\mu}}\partial{x_{\nu}}}} = \frac{\partial^{2}(F_{\mu\nu})_{0}}{{\partial{x_{\mu}}\partial{x_{\nu}}}}
\end{equation}
From this relation, we see that the total contribution of the {\it zero point field} arises from averaging over the vacuum fluctuations of the electromagnetic field up to the Compton scale. The peculiar nature of this relation is due to the domain of the Compton scale, where {\it zitterbewegung} effects are also present. Innumerable possibilities open due to the above consideration. It has been also shown by the author Sidharth \cite{bgs3} and others \cite{Rln} that the Dirac equation gets modified due to the non-commutative nature of space-time. This modification provides a remarkable explanation for the {\it Lamb shift} \cite{bgs4} in the energy levels of the hydrogen atom. Hence, we can perceive that the non-commutative nature of space-time, the Compton scale and the {\it zero point field} are very significant in the sense that they can explain several phenomena that are inexplicable by conventional methods.\\ \\


\begin{thebibliography}{99}
\bibitem{bgs} Sidharth, B.G. (2008). \emph{The Thermodynamic Universe}, World Sientific, Singapore 2008.
\bibitem{Battisti} Battisti, M.V. and Meljanac, S. (2009). \emph{Phys. Rev. D} 79, 067505 (2009); M.V. Battisti and S. Meljanac, \emph{Phys. Rev. D} 82, 024028 (2010).
\bibitem{Meljanac} Meljanac, S., Meljanac, D., Samsarov, A and Stojic, M. (2010).  \emph{Mod. Phys. Lett. A} 25, 579 (2010).
\bibitem{Chaichian} Chaichian, M.,  Sheikh-Jabbari, M.M., and Tureanu, A. (2001). \emph{Phys. Rev. Lett.} 86, 2716 – Published 26 March 2001.
\bibitem{Mendes} Vilela Mendes, R. (1996). \emph{Quantum mechanics and non-commutative space-time}, \emph{Physics Letters A}, 210 (1996) 232-240.
\bibitem{Camelia} Giovanni Amelino-Camelia and Valerio Astuti. (2015). \emph{Misleading inferences from discretization of empty spacetime: Snyder-noncommutativity case study}, \emph{Int. J. Mod. Phys. D} 24, 1550073 (2015).
\bibitem{Snyder}  Snyder, H.S. (1947).  \emph{Phys. Rev.} 71, 38-41 (1947); \emph{Phys. Rev.} 72, 68-71 (1947).
\bibitem {a} Sieberg and Witten, E. (2000). \emph{JHEP} 9 (2000) 032..
\bibitem {b} Hayakawa, M. (2000). \emph{Phys.Lett.B} 478 (2000) 394..
\bibitem {c} Sidharth, B.G. (2002). \emph{Foundation of Physics Letters} 15(5),2002,501ff.
\bibitem {d} Sidharth, B.G. (2005). \emph{The Universe of Fluctuations} (Springer, Netherlands).
\bibitem {db} Sidharth, B.G. (2001). \emph{Fuzzy, non commutative spacetime: A new paradigm for a new century} in \emph{Proceedings of Fourth International
Symposium on ``Frontiers of Fundamental Physics''} (Kluwer Academic/Plenum Publishers, New York), p.97--108.
\bibitem {e} Barut, A.O. and Bracken, A.J. (1981). \emph{Zitterbewegung and the internal geometry of the electron} \emph{Physical Review D}
Vol.23, No.10, 15 May 1981, pp.2454-2463.
\bibitem {diracpqm} Dirac, P.A.M. (1958). \emph{The Principles of Quantum Mechanics} (Clarendon Press, Oxford), pp.4ff, pp.253ff.
\bibitem {wigner} Salecker, H. and Wigner, E.P. (1958). \emph{Quantum Limitations of the Measurement of Space-Time Distances}
\emph{Physical Review} Vol.109, No.2, January 15 1958, pp.571-577.
\bibitem {bd} Bjorken, J.D. and Drell, S.D. (1964). \emph{Relativistic Quantum Mechanics}
(Mc-Graw Hill, New York), pp.58-59.
\bibitem {calphys} Calphys Institute:\emph{Introduction to Zero-Point Energy} \emph{http://www.calphysics.org/zpe.html}.
\bibitem {ijmpa} Sidharth, B.G. (1998). \emph{Int.J. of Mod.Phys.A} 13, (15), pp.2599ff.
\bibitem{bgs2} B.G. Sidharth, B.G., Abhishek Das and Arka Dev Roy. (2015). \emph{The Anomalous Gyromagnetic Ratio}, \emph{IJTP}, (DOI) 10.1007/s10773-015-2718-8.
\bibitem{bgs3}  Sidharth, B.G. (2010).  \emph{Noncommutative Spacetime, Mass Generation and Other Effects}, \emph{IJMPE}, Vol. 19, no. 1, pp. 79-90, 2010.
\bibitem {Rln} Raoelina, A. and Christian, R. (2011).  \emph{A Study of the Dirac-Sidharth Equation}, \emph{EJTP} 8, No.25, 177-182, 2011.
\bibitem{bgs4} Sidharth, B.G. and Abhishek Das. \emph{Revisiting the Lamb Shift}, to appear in \emph{EJTP}.
\end{thebibliography}
\end{document}